\documentclass[preprint]{aastex}
\usepackage{epsf}
\usepackage{natbib}
\usepackage{subfigure}
\usepackage{longtable, rotating}

\usepackage{epstopdf}

\newcommand{\degamin}{\mbox{$\hbox{$.\!\!'$}$}}
\newcommand{\degasec}{\mbox{$\hbox{$.\!\!"$}$}}
\newcommand{\feh}{\hbox{$ [{\rm Fe}/{\rm H}]$ }} 
\newcommand{\fehc}{\hbox{$ [{\rm Fe}/{\rm H}]$}} 
\newcommand{\dellpc}{\hbox{$ {\rm \Delta logP}$}}

\begin{document}

\title{The Determination Of Reddening From Intrinsic $VR$ Colors Of RR Lyrae Stars}

\author{Andrea Kunder and Brian Chaboyer} 
\affil{Dartmouth College, 6127 Wilder Lab, Hanover, NH 03755}
\affil{E-mail: Andrea.M.Kunder@dartmouth.edu and Brian.Chaboyer@dartmouth.edu}
\author{Andrew Layden} 
\affil{Bowling Green State University, Bowling Green, OH 43403}
\affil{E-mail: layden@baade.bgsu.edu}

\begin{abstract}
New $R$-band observations of 21 local field RR Lyrae variable stars 
are used to explore the reliability of minimum light $(V-R)$ colors as a tool
for measuring interstellar reddening.  For each star, $R$-band intensity mean 
magnitudes and light amplitudes are presented.  Corresponding $V$-band 
light curves from the literature are supplemented with the new photometry, 
and $(V-R)$ colors at minimum light are determined for a subset of these stars
as well as for other stars in the literature.  Two different definitions of
minimum light color are examined, one which uses a Fourier decomposition
to the $V$ and $R$ light curves to find $(V-R)$ at minimum $V$-band light,
$(V-R)_{min}^F$, and the other which uses the average color between
the phase interval 0.5-0.8, $(V-R)_{min}^{\phi(0.5-0.8)}$.
From 31 stars with a wide range of metallicities and pulsation periods, 
the mean dereddened RR Lyrae color at minimum light is
$(V-R)_{min,0}^F$ = 0.28 $\pm$ 0.02 mag and 
$(V-R)_{min,0}^{\phi(0.5-0.8)}$ = 0.27 $\pm$ 0.02 mag.  As was found by 
\citet{guldenschuh05} using $(V-I)$ colors, any dependence of 
the star's minimum light color on metallicity or pulsation amplitude 
is too weak to be formally detected.  We find that the intrinsic 
$(V-R)$ of Galactic bulge RR Lyrae stars are similar to those
found by their local counterparts and hence that Bulge RR0 Lyrae stars 
do not have anomalous colors as compared to the local RR Lyrae stars.  
\end{abstract}

\section{Introduction}
RR Lyrae variables play an important role in the determination 
of distances to the Galactic center, globular clusters, and 
neighboring galaxies.  This information is of
fundamental importance in the study of galactic structure and
dynamics.  In addition, RR Lyraes provide an excellent sample 
of stars to investigate kinematics in the galaxy (see, e.g., Beers 
et al. 2000, Layden et al.\ 1996, Dambis \& Rastorguev 2001).
Their pulsational properties combined with an almost constant 
absolute magnitude make them 
excellent tracers of the old stellar populations.  

As bulge RR Lyrae variables are among the oldest and most 
metal poor stars in the bulge, 
their distribution and signature are of considerable importance 
in determing the mix of populations in the Galactic bulge, and 
vital to our understanding of the nature of the bulge itself.  
RR Lyrae variables have also advanced our understanding especially 
of the halo and thick disk components of the Milky Way.
For example, the SDSS \citep{ive00} and QUEST \citep{vivas06}
surveys have used RR Lyrae 
stars to find complex substructure in the Galactic halo, thick disk 
and thin disk distribution.  Many irregular structures, such as 
the Sgr dwarf tidal stream in the halo and the Monoceros stream 
closer to the Galactic plane, have strengthened the notion
that the Milky Way is a complex and dynamical structure that is still being 
shaped by the merging of neighboring smaller galaxies.

Regardless of how convenient a distance indicator may be, with out
knowing reddening, accurate distances can not be found.
Especially the central regions of the Galaxy and other low
latitude areas suffer from severe crowding and high, patchy, reddening.
It would be especially useful to have a method of determine
interstellar reddening directly from the RR Lyrae, rather than 
depending on extinction maps in a certain area of the sky with
perhaps low resolution and large uncertainties.

The minimum light color of RR0 Lyrae variables have been used 
to estimate their line-of-sight reddenings.  This is based on a 
concept originally developed by \citet{sturch66} to investigate 
$\rm E(B - V)$, and refined by \citet{blanco92}.  \citet{blanco92} 
used 22 stars to derive a relationship between an RR Lyrae
variable's line of sight color excess, $\rm E(B-V)$, as a function of the RR 
Lyrae's period, metalicity and minimum $\rm (B-V)$ color.  
An investigation of $\rm E(V-I)$ was performed by \citet{mateo95} 
with an 11 star sample and expanded on by \citet{guldenschuh05} 
who used 16 RR0 Lyrae variables.  They found that $\rm E(V-I)$ depends
upon only minimum $\rm (V-I)$ color.  \citet{guldenschuh05} concluded that the 
intrinsic minimum light color of RRab variables is 
$\rm (V - I)_{min,0}^{\phi(0.5-0.8)}$ = 0.58  $\pm$ 0.02, with very 
little dependence on period or metallicity for periods 
between 0.39 and 0.7 days and metallicities in the range $\rm -3 \leq 
\feh \leq$ 0.  These studies define minimum light color to be the
average color over the phase range 0.5-0.8, designated here as
\textit{i.e.} $\rm (V - I)_{min}^{\phi(0.5-0.8)}$ for the color $\rm (V - I)$.  

Most recently, \citet{kunder08} did a study on the determination 
of $\rm E(V-R)$ from RR0 Lyrae stars minimum light color using 
11 stars and data in the literature.  They defined
minimum light color not by $\rm (V - R)_{min}^{\phi(0.5-0.8)}$, 
but by the $\rm (V-R)$ at minimum $V$-band light, 
$\rm (V-R)_{min}^F$.   Fourier fits to both the $V$- and $R$-band 
data were administered to determine this quantity.  This definition 
of minimum light was used because \citet{kanbur05} observed that 
the RR0 Lyrae in the LMC display period-color and amplitude-color 
relations that are flat at $\rm (V-R)_{min}^F$.   \citet{kunder08} found 
that $\rm (V-R)_{min}^F$ is a better diagnostic for determining 
intrinsic colors than $\rm (V-R)_{min}^{\phi(0.5-0.8)}$, because it is not 
as correlated with the RR Lyrae's amplitude.  \citet{kunder08}
find an intrinsic $\rm (V-R)_{min}^F$ to be 0.28 $\pm$ 0.014 mag for
periods between 0.39 and 0.6 days and metallicities in the range 
$\rm -1.7 \leq \feh \leq -$0.01.  

In this paper we report on new data gathered to refine the relationship 
of reddening at minimum light $\rm (V-R)$ color.  
Although $V$-band photometry of RR Lyrae stars is quite common in the
literature, this is not the case for the $R$-band.  Here
new $R$-band observations are obtained to triple the sample of RR Lyrae
stars available for intrinsic $\rm (V-R)_{min}$ determinations.  
Our 21 program stars were chosen to span a large range of 
$V$-amplitudes, pulsational periods and metallicities.
The $V$-band photometry for these stars is taken from 
various sources in the literature, including \citet{piersimoni93},
\citet{stepien72}, \citet{jones87}, \citet{bookmeyer77},
the Behlen Variable Star Survey\citep{schmidt91, schmidt95},
and the  All Sky Automated Survey (ASAS) 
database \citep{poimanski02}.  Both methods of determining 
minimum light are investigaed, $\rm (V-R)$ at minimum 
$V$-band light, $\rm (V-R)_{min}^F$, and the average color 
over the phase range 0.5-0.8, $\rm (V-R)_{min}^{\phi(0.5-0.8)}$.  
This reddening procedure can be applied to the thousands of 
RR Lyrae stars found in the MACHO database.

The MACHO Project was designed to search
for MACHOs through gravitational microlensing.  Their project
surveyed the same fields in the sky of the Large and 
Small Magellanic Clouds and the Bulge of the Milky Way.  Now,
after its eight years of operation from 1992-1999, the MACHO 
photometric database 
is an unprecedented resource for the study of stellar 
variability.  \citet{kunder08} found 3674 RR0 Lyrae stars
in the MACHO database, that were imaged simultaneously 
by the MACHO team in the $b_M$ and $r_M$ filters.  These
have been transformed to standard JohnsonÕs $V$ 
and Kron-Cousins $R$ \citep{alcock99}.
Using these RR Lyrae as standard candles, they can be used to map
the bulge and obtain its 3D structure, giving clues and constraining
models on the formation of the Milky Way Galaxy.
As with any study of the Galactic bulge, interstellar reddening 
estimates are essential to begin investigations in a quantitative
manner.

Recently there has been some controversy as to if RR Lyrae 
variables can be used to measure reddening from their
minimum light colors and if their absolute magnitudes are 
environment dependent (\textit{i.e.}, if the intrinsic properties of RR Lyrae
vary with the composition and age of the system/environment in which they 
reside).  Collinge, Sumi \& Fabrycky (2006) 
compared the RR Lyrae intrinsic color at minimum light, 
$\rm (V-I)_{min,0}^{\phi(0.5-0.8)}$, of OGLE Bulge 
RR Lyrae stars, dereddened according to \citet{sumi04}, with field 
RR Lyrae colors.  They concluded that there is a discrepancy of 
0.05-0.08 mag between the RR Lyrae-to-red clump (RC) color 
differential of the bulge population (measured from OGLE data) 
and that of the local population.  Whether this is a result of a RR Lyrae
or RC color discrepancy as a function of environment has
substantial implications on the use of absolute magnitudes of
RR Lyrae variables.  We apply the minimum light RR Lyrae 
color reddenings to the MACHO bulge RR Lyrae to check if the 
absolute magnitude of RR Lyrae stars are indeed 
environment-independent.  

The structure of this paper is the following.  The observations and 
data is given in \S 2.  The light curves and mean observational 
properties of the program RR Lyrae variables 
are presented in \S 3, and comments about individual stars
follow in \S 4.  A thorough investigation on the use of minimum 
light colors to calibrate $\rm E(V-R)$ is presented in \S 5.  In \S 6 
the intrinsic minimum light colors are then compared to RR Lyrae 
variables in the Galactic bulge and anomalous colors of Galactic 
Bulge RR Lyrae are discussed. 

\section{Observations and Reductions}
Observations were made over 31 nights at the WIYN 0.9 m telescope at Kitt 
Peak (on seven nights over 2006 May 4-11), and at the 1.3 m McGraw-Hill 
Telescope at the MDM observatory (on 24 nights
from 2006 September 29 - July 02, August 30 - September 7 and from 2008 
March 8-18).  The WIYN Telescope was equipped with the S2KB 2048 x 2048 CCD,
giving a field of view 20\degamin5 on a side, with a scale of 
0\degasec60 pixel$^{-1}$.  The 2048x2048 Echelle and 1024x1024 Templeton CCDs 
were used on the MDM telescope, giving fields of view of 17\degamin3 and 
8\degamin7 on a side, respectively, with scales of 0\degasec508 pixel$^{-1}$.
The observations were obtained primarily in the $R$- filter, but some $V$-band
images were also obtained.  On each night, twilight and
dawn flats were taken in the $R$- and $V$- filters.  The images were 
overscan-corrected, trimmed, and bias-subrated in the normal fashion using 
standard IRAF tasks.  

\citet{landolt92} standard stars were observed in the $R$- and $V$-bands
on three photometric nights, May 6, May 7 and May 8, 2006, 
with a wide range of color, air mass, and universal time that encompassed
the various properties of the program stars.  Ten stars in the vicinity of each 
RR Lyrae variable were selected as comparison stars, which could potentially 
be used to perform differential photometry on nights that were not photometric.
Instrumental magnitudes of the RR Lyrae variable, the comparison
stars and the standard stars were computed from each image using the 
DAOPHOT task PHOT \citep{stetson94}, as implemented in IRAF.

For each photometric night, transformation equations of the form
\begin{equation}
\label{Rtrans}
r-R = c_{R,0} + c_{R,1}X_{R} + c_{R,2}(V-R)
\end{equation}
and
\begin{equation}
v-V = c_{V,0} + c_{V,1}X_{V} + c_{V,2}(V-R)
\end{equation}
were constructed, where $r$ represents the instrumental magnitude, $R$ and $V$
is the standard magnitude from \citet{landolt92}, and $X_R$ and $X_V$ is the 
air mass in $R$ and $V$, respectively.  The coefficients were first obtained 
using least-squares minimization.  The coefficients $c_{R,2}$, the $R$-band 
color terms, for all three nights were averaged together to find
$<c_{R,2}>$.  The value of $<c_{R,2}>$ is then used as a constant in 
Equation~\ref{Rtrans} instead of $c_{R,2}$, and the coefficients 
$c_{R,0}$ and $c_{R,1}$ are found again.  This procedure was also carried out 
with the $V$-band color terms to find $<c_{V,2}>$, and to solve for 
$c_{V,o}$ and $c_{V,1}$.  

For the photometric night May 8, 2006, the transformation equation
\begin{equation}
\label{shorty}
r-R = c_{o} + c_{1}X_{R}
\end{equation}
was also used.  This is because for that particular photometric night, not 
many $V$-band images of RR Lyrae variables were collected and hence, for 
a subset of the RR Lyrae variables observed on this night, no $(V-R)$ 
information at a particular phase was known.  As the color terms found 
were small, $\rm <c_{R,2}>$ = $-$0.022 and $\rm <c_{V,2}>$ = $-$0.011,
and as the RR Lyrae colors do not vary much, the omission of the color 
term is small.  The rms scatter of the points around each best fit is
shown in Table~\ref{tab1}.  The number of stars used in each fit and the 
number of standard star fields observed that night is also shown.  
The rms from Equation~\ref{shorty} with no color term is indicated by $rms_{R,nc}$.
The number of calibration images for each RR Lyrae that uses a color term
($N_{VR}$) and the number of calibrations with no color term ($N_{R}$)
is listed for each star in Table~\ref{RRmags}.

Once these equations were determined, we applied them to all of the 
comparison stars. Every variable star field was observed at least twice on 
at least two different photometric nights.
Most variable star fields were observed 4-9 times over the three 
photometric nights.  When averaged together, the magnitudes for each 
comparison star agreed well, with the dispersion about the mean being less 
than 0.012 mag in most cases.  When performing differential
photometry, only comparison stars with dispersions about the mean being less 
than 0.025 mag were used.  On average, each RR Lyrae field had 3 suitable 
comparison stars.

Over the course of the data reduction, ACL noticed that the 
S2KB CCD encountered an excess of light falling near the center of the chip 
in the flat fields.  Thus a 2-4\% overillumination near the central region 
of the chip occured.  Using Layden's aperture mask, correction images were
developed that were applied to the data taken using this CCD.

The tables in Appendix~A present the photometry thus obtained. The columns 
contain (1) the Heliocentric Julian Date of midobservation (minus 2,450,000 
days), (2) the Johnson $R$ magnitude and (3) its error.

\section{Light Curves}
Almost all of the observed data were folded by the period listed in the 
General Catalogue of Variable Stars (Kholopov 1985).  For the stars SS CVn 
and AX Leo, the period from \citet{maintz05} was used, for UZ CVn the period 
from \citet{vandenbroere01} was used, and for IO Lyr and AN Ser, the period 
from \citet{leborgne07} was used.  
These periods yielded light curves in which our 2006 data and 2008 $R$-band
data were more closely aligned in phase space.  
As suggested by previous investigators, all twenty-one stars 
pulsate in the fundamental mode, and the stars AB UMa, RV UMa, and SS CVn, 
exhibit light curve modulation, the so called Blazhko 
effect.  There are other cases of stars with noticeable light curve scatter,
namely UY Boo, AX Leo, and BT Dra.  UY Boo has been shown by \citet{leborgne07}
to have irregular slow variations, and AX Leo has been reported by 
\citet{schmidt02} to show noticeable light curve scatter.  
\citet{piersimoni93} found that BT Dra has a bump that precedes the light 
minimum, and possibly, a hump on the rising branch.  We see light curve 
scatter at maximum and minimum light but no hump on the rising
branch for BT Dra.

\begin{figure}[htb]
\centering
\begin{tabular}{c c} 
\includegraphics[width=8cm]{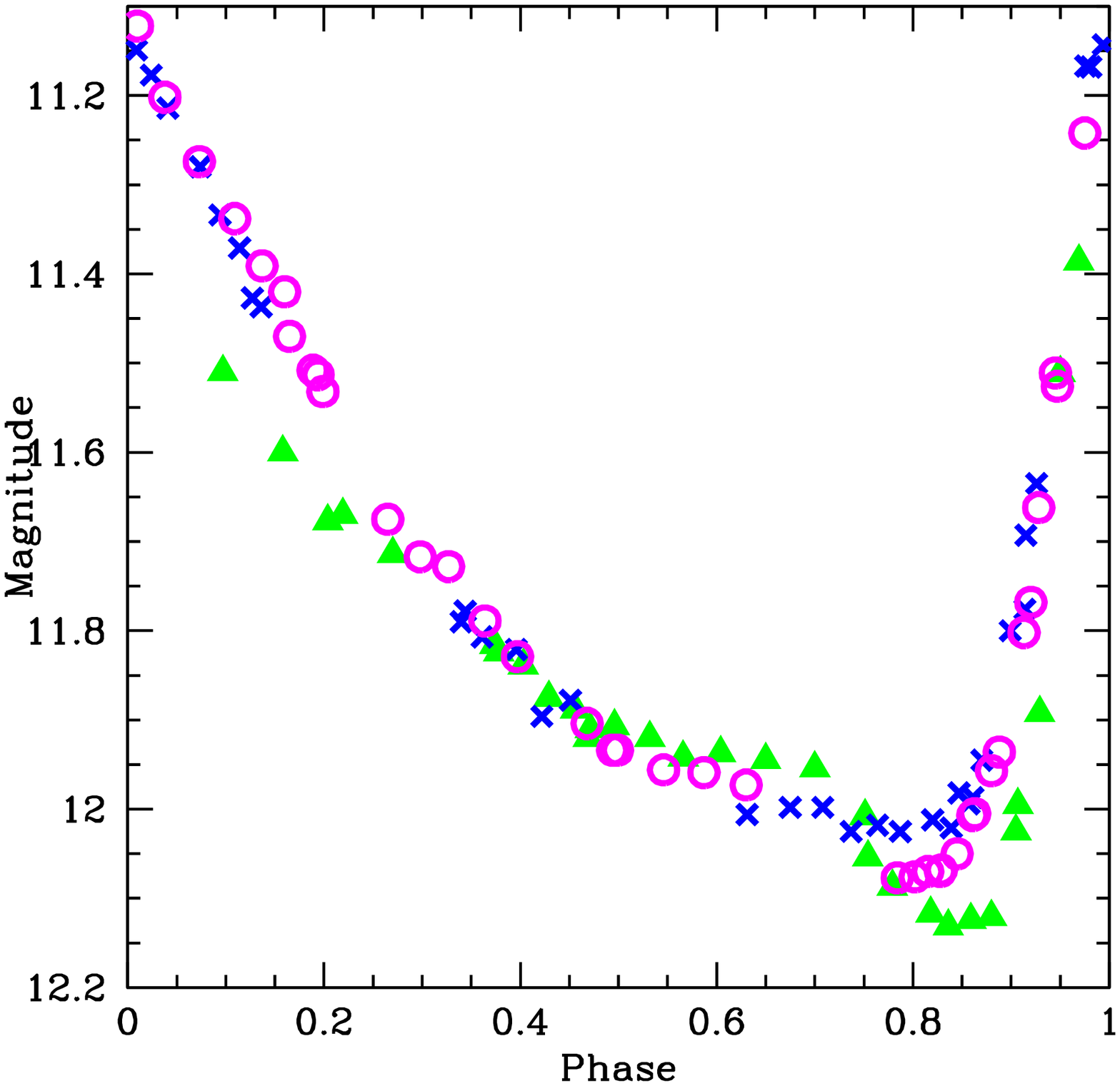} &
\includegraphics[width=8cm]{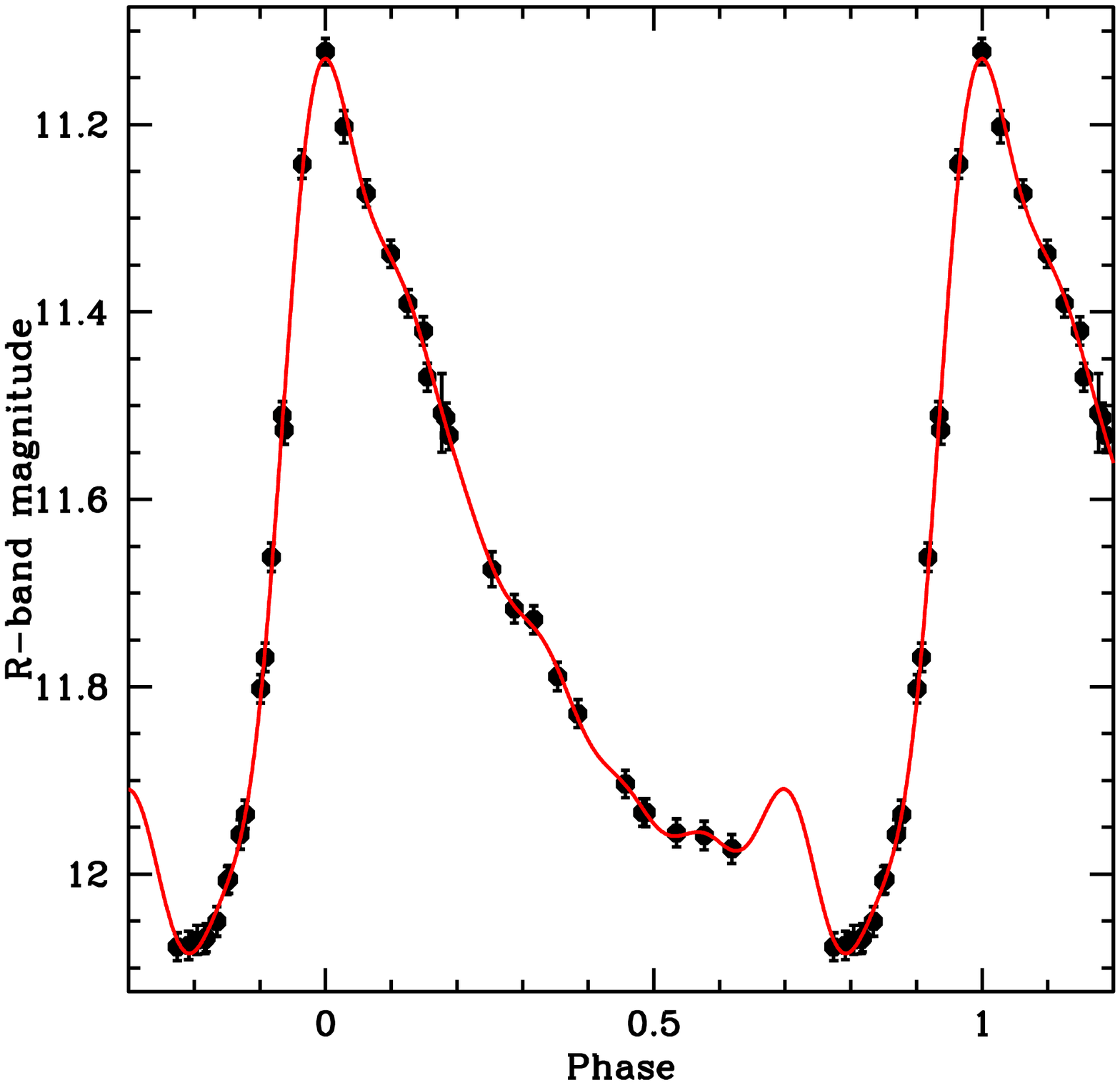} \\
\end{tabular}
\caption{$R$-band light curve of the Blazhko RR Lyrae SS CVn.  
{\textit Left:} The triangles represent data taken in 2006;
the crosses and circles represent data taken in 2008, at the 
beginning and end of the run, respectively. {\textit Right:} 
$R$-band light curve of SS CVn with the HJD range of 
4535.6571 - 4543.9312 days.  Over-plotted is the 8th order 
Fourier fit to the observed data.
\label{fig1}}
\end{figure}

\begin{figure}[htb]
\includegraphics[width=16cm]{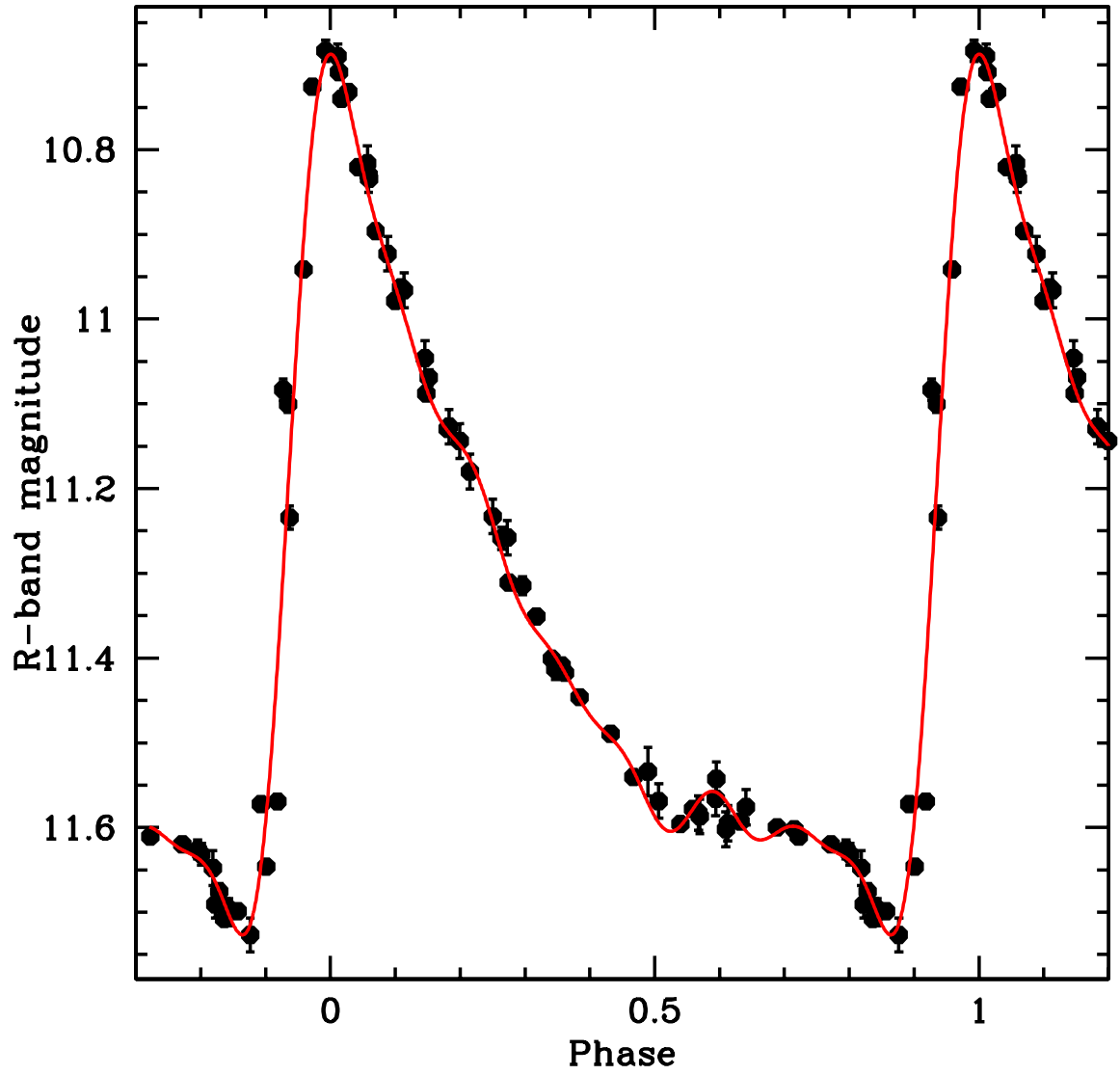}
\caption{$R$-band light curve of the RR Lyrae VZ Her.  
Over-plotted is the 8th order Fourier fit to the observed data.
\label{fig2}}
\end{figure}

The estimated intensity mean $R$-band magnitude and pulsation amplitudes 
for each star from the observed data is found from a Fourier decomposition 
of the light curve.  The order of the Fourier fit (from a 4th- to 8th-order) 
did not change the computed mean magnitude of the star.  However, variations 
of the amplitude, on the order of $\sim$0.01 mag, occurred depending on 
the order of the fit.  The error in the $R$-band amplitude is hence 0.01 
mag.  These values are presented in Table~\ref{RRmags} along with the number 
of observed data points for each star, $N_{obs}$, and the order of the 
Fourier fit used to determine the amplitude.  For the stars exhibiting 
scatter in their light curves due to the \textit{i.e.}, the Blazhko effect, a narrow range 
of HJD is used in determining the stars $R$-band magnitude and amplitude.  
For these cases, the limited HJD range shows no detectable scatter in
the light curve, yet there is ample data for an accurate
Fourier fit.  Individual light curves for two representative stars, one
Blazhko star and one normal star, are seen in Figures~\ref{fig1}-\ref{fig2}.

Corresponding $V$-band light curves from the literature are found for
the program stars.  This is so that the $\rm (V-R)$ at minimum $V$-band
light, $\rm (V-R)_{min}$, can be calculated for each of the program stars.  An
analysis on the use of $\rm (V-R)_{min}$ for reddening 
determination can thus be carried out.

Almost half of the corresponding $V$-band light curves in 
our sample are obtained from the All Sky Automated Survey (ASAS) 
database \citep{poimanski02}.  The ASAS data used
covers a range of observations from 2001-2009
and for each measurement, five $V$-band magnitude values are 
given which correspond to five different  aperture sizes used in 
the photometry.   The time series, $V(t)$, was generated by 
weighting the average of the magnitude values given for 
the five apertures as described by \citet{kovacs05}.  
The corresponding $V$-band data for our program stars is 
mentioned in the individual notes for each RR Lyrae star.

The $V$ and $R$ light curves were fit with a series of six RR Lyrae
light curve templates as described in \citet{layden98} to find where 
maximum light occurs and align it with phi=0.  For stars with 
noticeable light curve scatter (that was not due to the Blazhko effect),  
the light curve was separated into different different observing seasons, 
which usually corresponded to different light curve ``branches".    
The period for each of these ``branches" was solved for,
and again aligned with phi=0.  Hence, the phase-shift is treated 
as a free parameter.  Since color and reddening is not related 
to the long-term period stability, this is an acceptable way to reduce 
light curve scatter, especially when combining photometry from 
widely differing years.

\section{Comments on Individual Stars}
O-C curves are widely used for investigating the period behavior of variable 
stars.  The O-C value refers to Ôobserved - calculated:Õ the difference 
between the observed and calculated epoch of a light curve.  

The GEOS RR Lyr Survey
\citep{leborgne07}, contains $\sim$50000 times of maximum light from more than 
3000 RR Lyr stars obtained either visually or with electronic devices or 
photographically.  From these measurements, O-C curves are made.

\textit{UY Boo}  This star has an O-C curve that changes by $\sim$4 days 
from approximately HJD 2,420,000 to 2,450,000 days.  \citet{leborgne07} shows 
that UY Boo exhibits irregular slow variation.  
We notice extreme light-curve scatter in the $R$-band data.
To find its photometric
properties, we use the UY Boo data with the HJD range indicated in 
Table~\ref{RRmags}.  The $V$-band taken for this star is taken from ASAS.  To find
$\rm (V-R)_{min}$, separate
results are presented for the 2006 and 2008 data to describe the star's behaviour
at these distinct points in its irregular variation. 

\textit{TV CrB}  This star is shown to have a well-defined linearly increasing 
period \citep{leborgne07}.  It has an O-C curve that changes in 
a relatively linear manner by $\sim$0.2 days from approximately 
HJD 2,410,000 to 2,450,000 days.  We see no light curve 
scatter in the $R$-band between the 2006 and 2008 data.  The  
$V$-band data is taken from both the Behlen Survey and 
Layden et al. (in preparation).  

\textit{SS CVn} This star exhibts the Blazhko effect with a Blazhko 
period of 97 days (Wils, Lloyd, \& Bernhard 2006).  Similarly, we find 
that the 2006 and 2008 $R$-band data do not match up well with each other.  
As their is no $V$-band data with approximately the same HJD range 
as the $R$-band data in the literature, no $(V-R)_{min}$ was found for
this star.

\textit{UZ CVn} This star has an O-C curve that  is shown to be represented either by an 
abrupt period change around epoch 28000 or by a parabolic fit 
\citep{vandenbroere01}.  From approximately HJD 2,430,000 to 2,450,000 days, 
the O-C curve changes by $\sim$0.4 days.  We see no evidence of light curve
variation in our 2006 and 2008 data.  The $R$-band data used 
is from the Behlen Survey combined with the new observations presented in 
this paper.  The $V$-band is taken from Behlen Variable
Star Survey.  We find
that the period given by \citet{vandenbroere01}, 0.69779191 days, yields not 
only a more tightly phased light curve in the Behlen data,
but also results in a better fit between the $R$-band Behlen data and the 
data presented here. 

\textit{BC Dra}  The $V$-band photometry is taken from \citet{szabados82}, \citet{szabados82}, 
with a mean $V$ error of 0.036 mag.  

\textit{BT Dra}  This star has an O-C curve with a roughly linear trend in 
decreasing O-C, with a 0.1 day variation from approximately HJD 2,420,000 
to 2,450,000 days.  The $V$-band data is taken from \citet{piersimoni93} who 
note that the light curve of BT Dra has a bump that precedes minimum light, 
and possibly, a bump on the rising branch.  We find scatter in the $R$-band data 
of $\sim$0.05 mag at maximum and minimum light.  This is not 
seen for the rest of the light curve.

\textit{SW Dra} This star has a 
significant phase-lag between light and radial velocity curves, and its O-C 
curve shows a gradual increase from approximately HJD 2,440,000 to 2,455,000 
days.  The $V$-band data is taken from both \citet{stepien72} and 
\citet{jones87}.  Due to few bright stars in its vicinity, only one 
comparison star was suitable to use for the $R$-band light curve.  

\textit{CW Her}  This stars' O-C curve changes by $\sim$0.2 days from 
approximately HJD 2,440,000 to 2,450,000 days.
The $V$-band light curve is taken from 
the Behlen Variable Star Survey.  

\textit{VZ Her} This star changes its period (Szeidl, Olah \& Mizser 1986).  
This $V$-band data is from Fitch, Wisniewski, \& Johnson (1966) and \citet{sturch66}.  

\textit{AX Leo}  This star shows noticeable light curve scatter \citep{schmidt02}.
From our $R$-band observations we find a 0.03 mag discrepancy at 
some points in the light curve between the 2006 and 2008 data.
The $V$- and $R$-band light curve is taken from the
Behlen Survey and only data from 1990 is used.  The data from the Behlen
Survey taken after 1990 is too sparse to accurately determine 
its' properties.

\textit{FN Lyr}  This star has the largest E(B-V) value in the sample, and hence
it is the most uncertain.  This
star was not used in the calibration of $\rm (V-R)_{min,0}$.

\textit{IO Lyr}  The $V$-band data from \citet{stepien72} and \citet{sturch66} 
has sparse phase coverage at minimum light (2 data points).  Hence this data is combined
with the $V$-band data from Layden et al. (in preparation).  No significant
difference in $\rm (V-R)_{min}$ is found when using only the \citet{stepien72} and \citet{sturch66} 
data, or when using the Layden et al. (in preparation) data.

\textit{AN Ser } The $V$-band light curve is taken from ASAS.  
The $V$-band photometry taken by \citet{lub77} and transformed to
Johnson $V$ agrees very well with the more current
ASAS photometry, although
they are taken more than almost 40 years apart.

\textit{AT Ser}  This star has an O-C curve that changes by $\sim$$-$0.6 days 
from approximately HJD 2,418,000 to 2,455,000 days.  It then seems to hover 
around zero.  There is no noticeable light curve discrepancies between the 
2006 and 2008 $R$-band data.  The $V$-band is taken from ASAS Project.

\textit{AW Ser}  The $V$-band data from ASAS project has a lot of scatter, 
especially at minimum light.  This could be due to its relatively faint 
magnitude. Its O-C curve has only 9 points, but does show a change
by $\sim$ 0.2 days from approximately HJD 2,448,000 to 2,455,000 days.  We
do not see any difference in the $R$-band 2006 and 2008 data.
The data from Layden et al. (in preparation), corresponds very 
well with the ASAS data, but shows significantly less light curve scatter.
Hence, the Layden data is used.

\textit{AB UMa} This star exhibits the Blazhko effect, and hence
there is a significant discrepancy in the $R$-band 2006 and 2008 light curve 
data.  In the determination of $\rm (V-R)_{min}$, both the $V$- and $R$-band 
data from the Behlen Survey is used, which is taken simultaneously, to 
avoid inconsistencies in the light curve due to different cycles in the 
Blazhko light curve.  Two separate
results are presented to describe the star's behaviour
at these distinct points in its Blazhko phase. 

\textit{RV UMa} This star exhibts the Blazhko effect with a Blazhko 
period of 93 days \citep{wils06}.  A thorough light curve analysis
of RV UMa by \citet{hurta08} reveals the quintuplet frequency solution of this 
star.  The $R$-band data shows clearly a discrepancy in the 2006 and
2008 data.  No $V$-band data with the same HJD range as that of the $R$-band
was found, and so no $\rm (V-R)_{min}$ was found for this star.

\textit{AT Vir} This star is shown to have a linearly decreasing 
period \citep{leborgne07}.  It shows a parabolic shape in its O-C diagram, 
varying by $-$0.4 days from approximately HJD 2,420,000 to 2,450,000 days.  
We do not find any $R$-band light curve  
scatter between the 2006 and 2008 data, although admittedly, there in not a
wealth of $R$-band data for this star.
The $V$-band is taken from ASAS Project.  

\textit{AV Vir} The $V$-band light curve is taken from the ASAS Project.  

\textit{ST Vir} \citet{leborgne07} find
that the O-C pattern for this star shows evidence for several changes in the 
direction of the period variation.  We do not see any scatter in our 2006 
and 2008 $R$-band data.  The $V$-band data is taken 
from both the ASAS Project.

\section{Reddening Calibration}
The apparent $\rm (V-R)$ colors of the program stars are
found at minimum $V$-band light by (1) performing a Fourier decomposition
on the $V$- and $R$-band light curves and finding the minimum $\rm (V-R)$
at minimum $V$-band light, $\rm (V-R)_{min}^F$, and (2) finding the average color
for phases between 0.5 and 0.8, $\rm (V-R)_{min}^{\phi(0.5-0.8)}$.  
\citet{kunder08} noticed a slight correlation 
between amplitude and $\rm (V-R)_{min,0}^{\phi(0.5-0.8)}$.  However, 
to obtain $\rm (V-R)_{min}^F$, a Fourier decomposition is performed
which requires a relatively 
complete light curve, especially at minimum light.  We investigate these
two methods further to asses how well the apparent $\rm (V-R)$ colors
of RR Lyrae stars can be used to find the interstellar reddening along the
line of sight to the star.

When fitting a Fourier decomposition to the $V$ and $R$-band data,
the same fit order for the $R$-band listed in 
Table~\ref{RRmags} is used; the fit order for the $V$-band is given in
Table~\ref{redval}.  In general, an 8th-order fit was used for both the 
$V$- and $R$-band light curves.  The apparent $\rm (V-R)$ color at minimum 
$V$-band light, $\rm (V-R)_{min}^F$, is dereddened from the reddening values
taken from Schlegel, Finkbeiner \& Davis (1998).  The adopted \rm $E(V-R)$ 
values are shown in Table~\ref{redval}, as well as the computed intrinsic 
color at minimum $V$-band light, $\rm (V-R)_{min,0}^F$ (column 7) and the error in the 
error in the color at minimum $V$-band light, $\rm \sigma (V-R)_{min,0}^F$.  
The $V$-band amplitudes, as determined from the $V$-band Fourier fits,
are also shown.  

To estimate an error in $\rm (V-R)_{min}^F$ value for our program RR Lyrae stars, 
various tests are performed to asses how much $\rm (V-R)_{min}^F$ changes.  
First, a few points in the $V$ and $R$ light curves were removed, particularly 
those points close to minimum $V$-band light.  A Fourier decomposition was 
applied again to these modified light curves, and $\rm (V-R)_{min}^F$ was 
re-calculated.  The $\rm (V-R)_{min}^F$ changed by $\sim$0.005 mag.  This 
indicates that the determination of $\rm (V-R)_{min}^F$ is largely independent on 
the inclusion or exclusion of a few data points in the light curve.  This is 
most likely due to the fact that most of the light curves presented here are quite 
complete.  There are some stars with not many points in their light curves, 
e.g., UY Boo, and the adopted for $\rm (V-R)_{min}^F$ is larger for these stars.  

Second, different fit-orders are employed.  In this case, the $\rm (V-R)_{min}^F$ 
value can change by $\sim$ 0.02 mag.  All fits are examined by eye to see 
which Fourier fit approximates the light curve best, especially the fit at 
minimum $V$-band light.  From these two tests, an error of 0.01 mag in 
$\rm (V-R)_{min}^F$ is estimated.  Stars with either noisy or sparse light 
curves have larger adopted errors.  The error in $\rm E(V-R)$ is also 
about 0.01 mag.  Hence, most of the adopted errors in our $\rm (V-R)_{min,0}^F$ 
values is 0.01 mag.  The error in $\rm (V-R)_{min,0}^F$ is given in Table~\ref{redval}.

The zero-point error in the $R$-band light curve influences 
both $\rm (V-R)_{min}^F$ and $\rm (V-R)_{min}^{\phi(0.5-0.8)}$.  
This varies from star to star, depending mainly 
on the sigma in the calibrated magnitude of the comparison stars 
and the number observations of the comparison star.  In general,
a comparison star has a sigma in the calibrated magnitude that 
is less than 0.021 mag, and most have a smaller sigma ($\sim$0.012 mag).  
Also, in general each field was observed photometrically 3 or more times
which results in a zero-point error that is 0.010 mag or less.  The two 
RR Lyrae fields with SW Dra and AX Leo have only 2 calibration 
observations, and have slightly larger zero-point uncertainties of 
$\sim$0.018 mag and $\sim$ 0.011 mag, respectively.  The zero-point 
error is taken into account when calculating the error in the individual 
light curve data points presented in Appendix~A and when performing 
the Fourier decomposition to the light curve.  

Computing the colors at minimum light via an arithmetic mean of the observed
$\rm (V-R)$ data having phases between 0.5 and 0.8 is relatively straight forward.
The mean values, in magnitude units, their standard errors of the mean, and
the number of points at minimum light in the $V$ and $R$ light curves
are reported in the last three columns of Table~\ref{redval}.

Three additional stars not observed in our project are included
Table~\ref{redval} and in our subsequent analysis.  
The two stars, AL CMi and GO Hya, have
photometry from the Behlen Variable Star Survey \citep{schmidt91,
schmidt95} and not only have adequate phase coverage for a good estimate 
of $\rm (V-R)_{min}$, but also well known reddening values.  
The third star not observed in our project is TZ Aur, and the
photometry is taken from \citet{warner08}.  

\begin{figure}[htb]
\includegraphics[width=16cm]{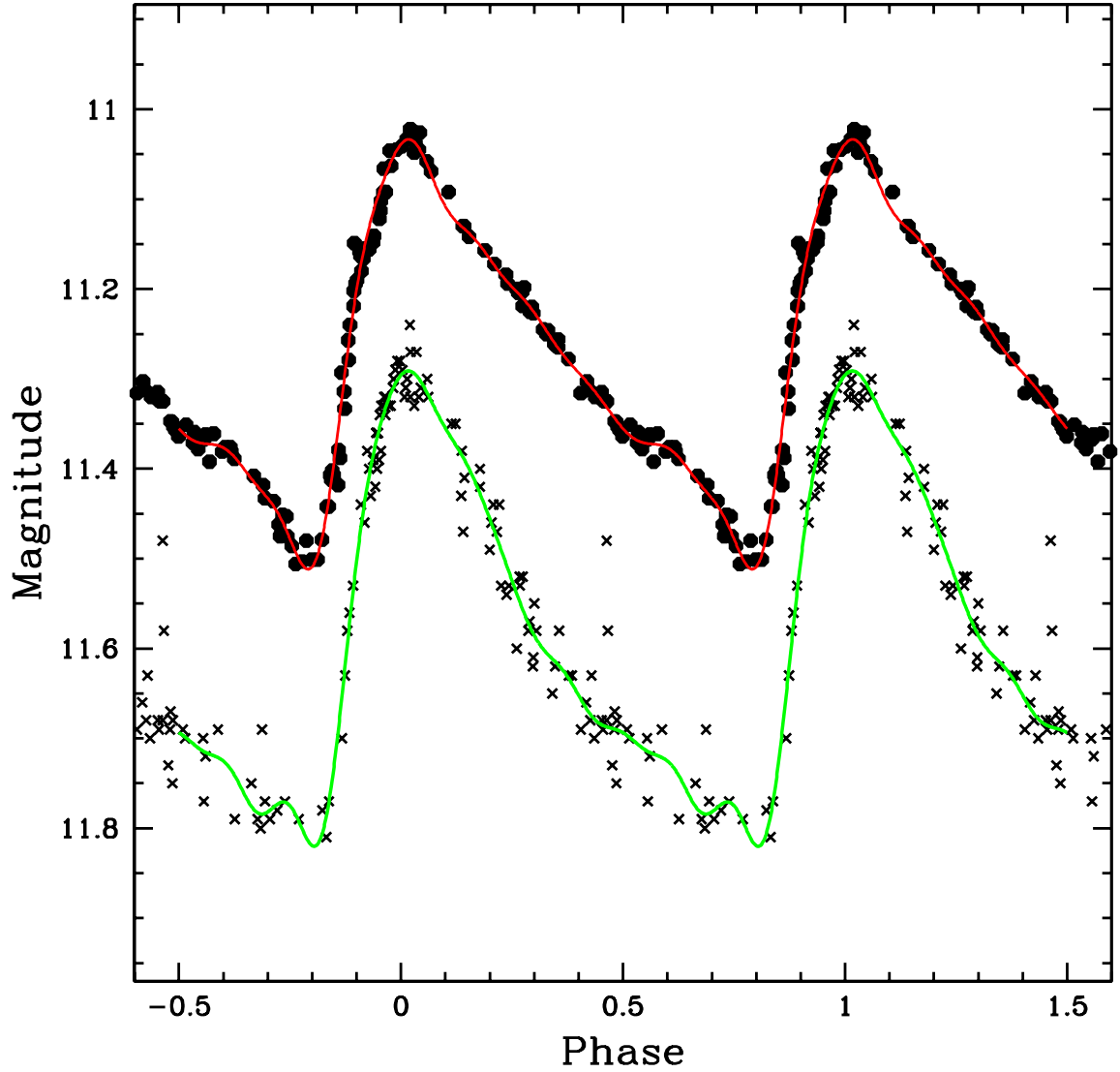}
\caption{$V$- and $R$-band light curves of the RR Lyrae star BC Dra.
The circles are from the $R$-band observations presented here, and 
the crosses are from $V$-band observations from \citet{szabados82}.
The solid lines designate an 8th order Fourier fit to the observed data.
\label{fig3}}
\end{figure}

\begin{figure}[htb]
\includegraphics[width=16cm]{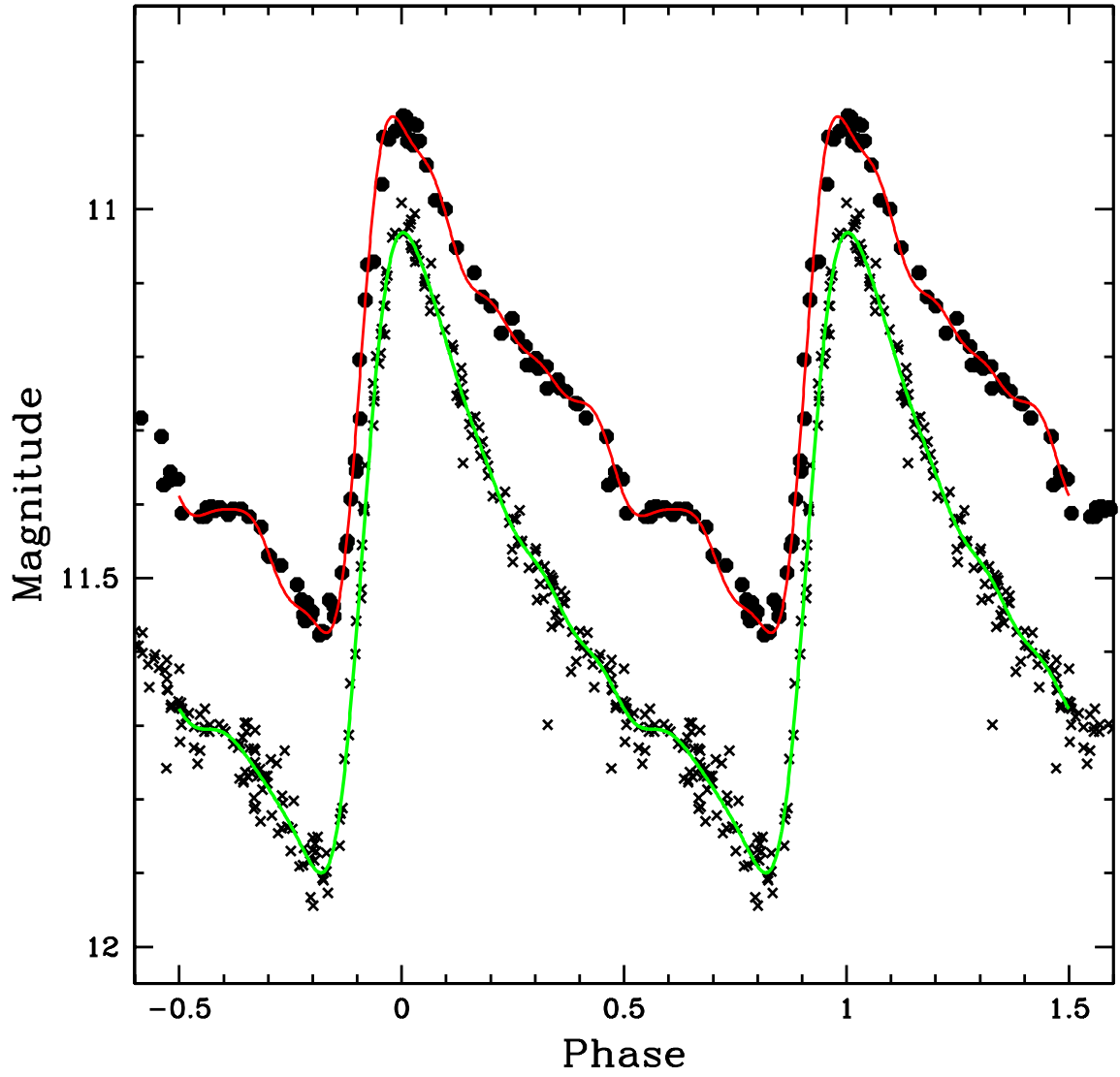}
\caption{$V$- and $R$-band light curves of the RR Lyrae star AT Ser.  
The symbols are as in Figure~\ref{fig3}, only the $V$-band data
is taken from the ASAS Project.
\label{fig4}}
\end{figure}

For three stars observed from our program, (AB UMa, AX Leo, CW Her) 
photometry from the Behlen Survey was used exclusively in determining 
$\rm (V-R)_{min}$.  This is because these
stars had light curves that exhibited light curve variation as a function of 
Julian Date, and none of these stars had $V$-band data with the
same HJD as that of the $R$-band data presented here.
The Behlen Survey data has the advantage that $V$ and $R$-band
are taken simultaneously.  The scatter in light curve due to the intrinsic
properties of the star, \textit{i.e.,} Blazhko effect, irregular slow variations, etc. 
is exhibited identically the light curves for both passbands.
For these stars in which the $V$- and $R$-band light curves
were taken simultaneously, an error of 0.01 mag in $\rm (V-R)_{min,0}^F$ 
is assigned.  For the star, UZ CVn the $R$-band data from the Behlen
Variable Star Survey is supplemented with that from this paper.
The combined $V$- and $R$- band photometry used to determine
$\rm (V-R)_{min}$ is shown for two
representative stars in Figures~\ref{fig3} and \ref{fig4}.

When performing a Fourier decomposition on the $V$- and $R$-band light
curves and finding the $\rm (V-R)$ at minimum $V$-band light, the mean 
$\rm (V-R)_{min,0}^F$ is 0.28 $\pm$ 0.02 mag, where 0.02
is the dispersion about the mean.  Finding the average $\rm (V-R)$ color for
phases between 0.5 and 0.8 results in a mean $\rm (V-R)_{min,0}^{\phi(0.5-0.8)}$ 
of 0.27 $\pm$ 0.02 mag.  These results are within the error bars of each other.

\begin{figure}[htb]
\includegraphics[width=16cm]{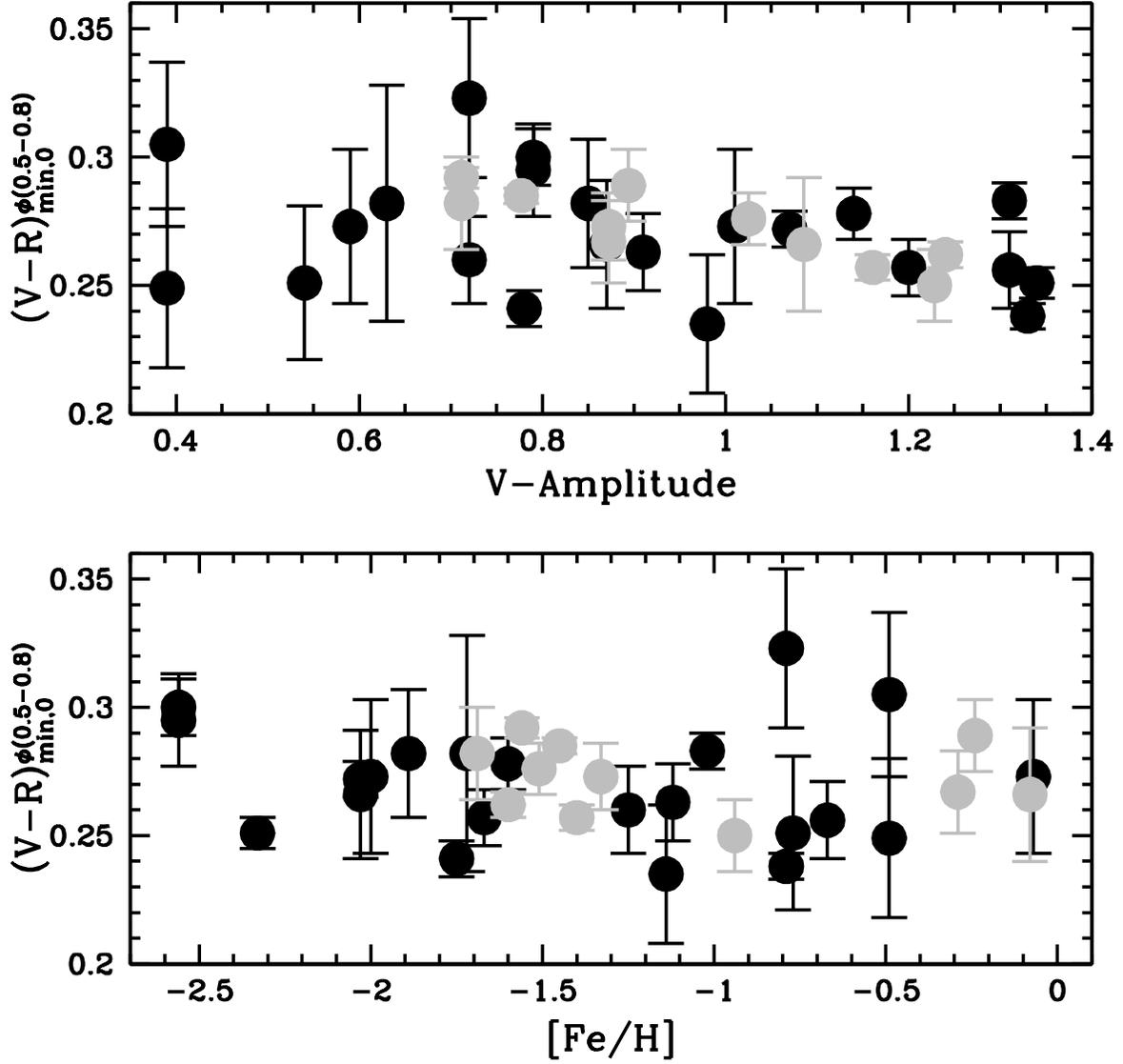}
\caption{The minimum $\rm (V-R)$ color,  $\rm (V-R)_{min,0}^{\phi(0.5-0.8)}$, 
is shown as a function of \feh and $V$-amplitude.  The grey circles represent
stars previously analyzed by \citet{kunder08}.
\label{fig5}}
\end{figure}

\begin{figure}[htb]
\includegraphics[width=16cm]{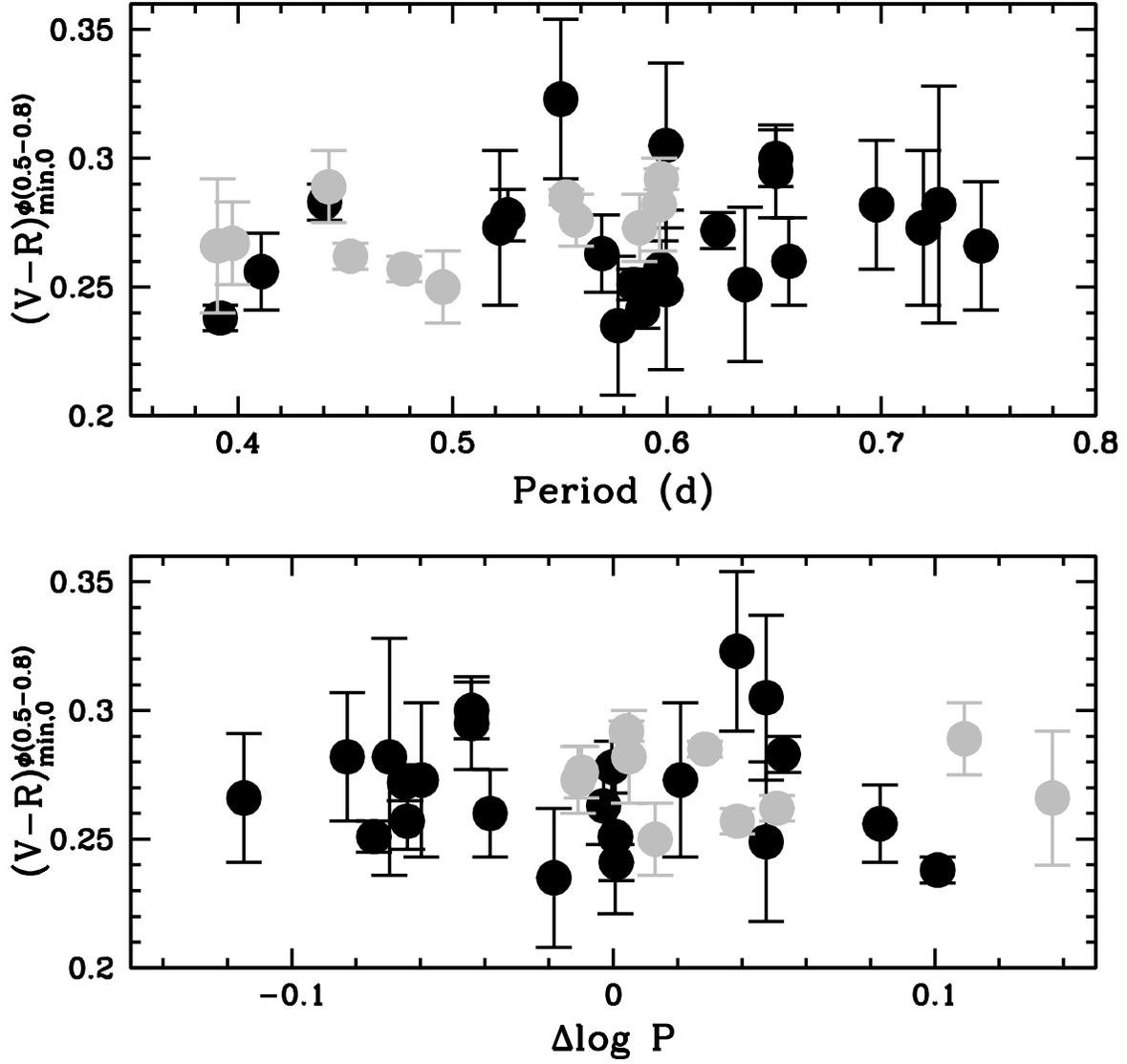}
\caption{The minimum $\rm (V-R)$ color, $\rm (V-R)_{min,0}^{\phi(0.5-0.8)}$, 
is shown as a function of period and \dellpc.  Symbols are the same
as in Figure~\ref{fig5}.
\label{fig6}}
\end{figure}

Figure~\ref{fig5} and Figure~\ref{fig6} show the intrinisic color 
at minimum $V$-band light, $\rm (V-R)_{min,0}^{\phi(0.5-0.8)}$,
as a function of period, \feh, $V$-band amplitude and \dellpc.
There is not much difference when using $\rm (V-R)_{min,0}^F$.
The period shift, \dellpc, is the difference between the periods of 
RR Lyrae variables at fixed amplitude.  An RR Lyrae amplitude is 
usually considered to be a function of temperature (see Clement \&
Shelton 1999, Kunder \& Chaboyer 2009).  This comparison is usually
made between the RR0 Lyrae stars in different globular clusters
\citep{sandage81, carney92, sandage93}, and is
thought to indicate the evolutionary state of an RR Lyrae variable.

The mean $\rm (V-R)_{min,0}^{\phi(0.5-0.8)}$ is independent 
or at least very insensitive to metallicity, amplitude and \dellpc.  
No trend is seen in a least squares analysis of 
$\rm (V-R)_{min,0}^{\phi(0.5-0.8)}$ as a function of \fehc, \dellpc, 
period or $V$-amplitude.  Weighted least squares fits led to
slopes of  +0.02 $\pm$ 0.04 mag $\rm day^{-1}$ (for period),
$-$0.023 $\pm$ 0.014 (for $V$-amplitude), +0.016 $\pm$ 0.05 
mag $\rm dex^{-1}$ (for \fehc), and +0.026 $\pm$ 0.02 mag 
$\rm day^{-1}$ (for $\dellpc)$.  These results are very similar 
for $\rm (V-R)_{min,0}^F$, when finding the $(V-R)$ at
minimum $V$-band light from a Fourier decomposition.  
Weighted least squares fits led to slopes of  +0.07 $\pm$ 0.04 
mag $\rm day^{-1}$ (for period), $-$0.031 $\pm$ 0.012 (for 
$V$-amplitude), +0.002 $\pm$ 0.005 mag $\rm dex^{-1}$
(for \fehc), and +0.0003 $\pm$ 0.06 mag $\rm day^{-1}$ 
(for $\dellpc)$. 

The greatest evidence of an $\rm (V-R)_{min,0}$ dependence is with $V$-amplitude,
where a 2-sigma slope in the $\rm (V-R)_{min,0}$ vs $V$-amplitude 
plane is found.  This is the case for both methods of finding minimum light.  Although
in both cases the slope with $V$-amplitude is insignificant, the 
slope with $\rm (V-R)_{min,0}^F$ is larger, in contrast with the 
finding of \citet{kunder08}.  The larger sample size shows that both
methods for determining minimum light are independent on $V$-amplitude.

Of particular interest is the relation of $\rm (V-R)_{min,0}$
with \fehc.  \citet{guldenschuh05} did not find a trend with 
minimum light $\rm (V-I)$ color and \fehc, and no trend is seen
with this larger of sample of stars either.  
It is interesting to note an increase of scatter in 
$\rm (V-R)_{min,0}$ in the more metal-rich stars.  This 
is evident in the \citet{blanco92} sample as well, and may 
be absent in the $\rm (V-I)_{min,0}$ sample due to the 
much smaller sample size.

If we seek a multidimensional fit of $\rm (V-R)_{min,0}$ with 
period, \feh, $\rm \fehc^2$, we obtain
\begin{equation}
\label{forcefit}
(V-R)_{min,0}^{\phi(0.5-0.8)} = (0.21 \pm 0.02) + (0.07 \pm 0.04)\ P - (0.048 \pm 0.016)\ \feh - (0.02 \pm 0.01)\fehc^2
\end{equation}
and
\begin{equation}
(V-R)_{min,0}^F = (0.27 \pm 0.02) +- (0.01 \pm 0.05)\ P - (0.01 \pm 0.02)\ \feh + (0.001 \pm 0.008)\fehc^2
\end{equation}
with an rms of 0.022 mag and 0.024 mag, respectively.  
None of these slopes are formally significant, although 
\citet{blanco92} determined $\rm E(B-V)$ 
using these parameters.  That the $\rm (V-R)_{min,0}^F$ is largely
independent on period, amplitude and \feh is the same conclusion 
reached by \citet{kunder08}, but the result here is based on 31 RR0
Lyrae stars, as opposed to eleven.

\section{Are the Colors of Galactic Bulge RR Lyrae Anomalous?}
Stutz, Popowski \& Gould (1999) found that the $\rm (V-I)_0$ colors 
of the bulge RR Lyrae stars behave in an anomalous way, distinct 
from the $\rm(V-I)_0$ colors of local stars.  For a fixed period, the Baade's 
window RR Lyrae stars are $\sim$0.17 mag redder in $\rm (V-I)_0$ than the 
local RR Lyrae stars.  
\citet{popowski00} 
showed that part of these offsets were due to errors in the original 
photometry used.  The other part, however, is unclear and various 
attempts to explain such an offset is reviewed by \citet{popowski00}.
\citet{paczynski98} and \citet{stutz99} explain this offset in terms of 
a difference between the intrinsic properties of the Red Clump and 
RR Lyrae stars in the two populations.  \citet{popowski00} noted 
that a non-standard interstellar extinction of $R_V$= 2.1, rather 
than the standard value of 2.5, would cause the $(V-I)_0$ 
discrepancy between bulge and local RR Lyrae disappear.  

Recently, \citet{collinge06} found a discrepancy of 0.05 mag between the 
mean value of $\rm (V-I)_{min,0}$ in the OGLE RR Lyrae sample in the Galactic bulge
and in the field RR0 Lyrae value obtained by \citet{guldenschuh05}.
The reddening values used in determining OGLE RR Lyrae $\rm (V-I)_{min,0}$ was taken
from the reddening map from \citet{sumi04}, which in turn uses
the observed color of the red clump (RC) giants as a measure
of reddening.  There are three possible causes for this discrepancy.\\

(1) The $\rm E(V-I)$ values reported by \citet{sumi04} are 0.05-0.08 mag
too large, most likely caused by a discrepancy between the
RR Lyrae-to-red clump color differential of the bulge
population (measured from the OGLE data)
and that of the local population.\\
(2) The \citet{guldenschuh05} result is calibrated incorrectly by 0.05-0.08 mag.
Their result is based on 16 calibrating stars.  \\
(3) The intrinsic colors or RR Lyrae at minimum light are a 
function of environment.  If this were the case, there should be
significant doubts on the validity of using
RR Lyrae stars as distance indicators.
The suggestion of an RR Lyrae intrinsic color discrepancy between the bulge and 
the local region of the Galaxy has been made before \citep{stutz99, popowski00}.  
An age discrepancy between RR Lyrae in the Bulge and in the field could 
perhaps be the source of such an intrinsic color discrepancy, such as seen
in the red clump stars
\citep{udalski98, girardi01}.  It may be that population effects are more important 
for the red clump than for the RR Lyrae, but further studies of local RR Lyrae colors are 
needed before this can become a strong statement. 

We take a thorough look at this possibility, if the intrinsic colors 
or RR Lyrae at minimum light are a function of environment.  
The RR0 Lyrae stars from the MACHO Project \citep{kundchab08} 
are used for this purpose.  As the MACHO light curves have a wealth 
of data points and good phase coverage, the $\rm (V-R)$
at minimum light $V$-band light from the Fourier decomposition is used 
to find $\rm (V-R)_{min}$.

\begin{figure}[htb]
\includegraphics[width=16cm]{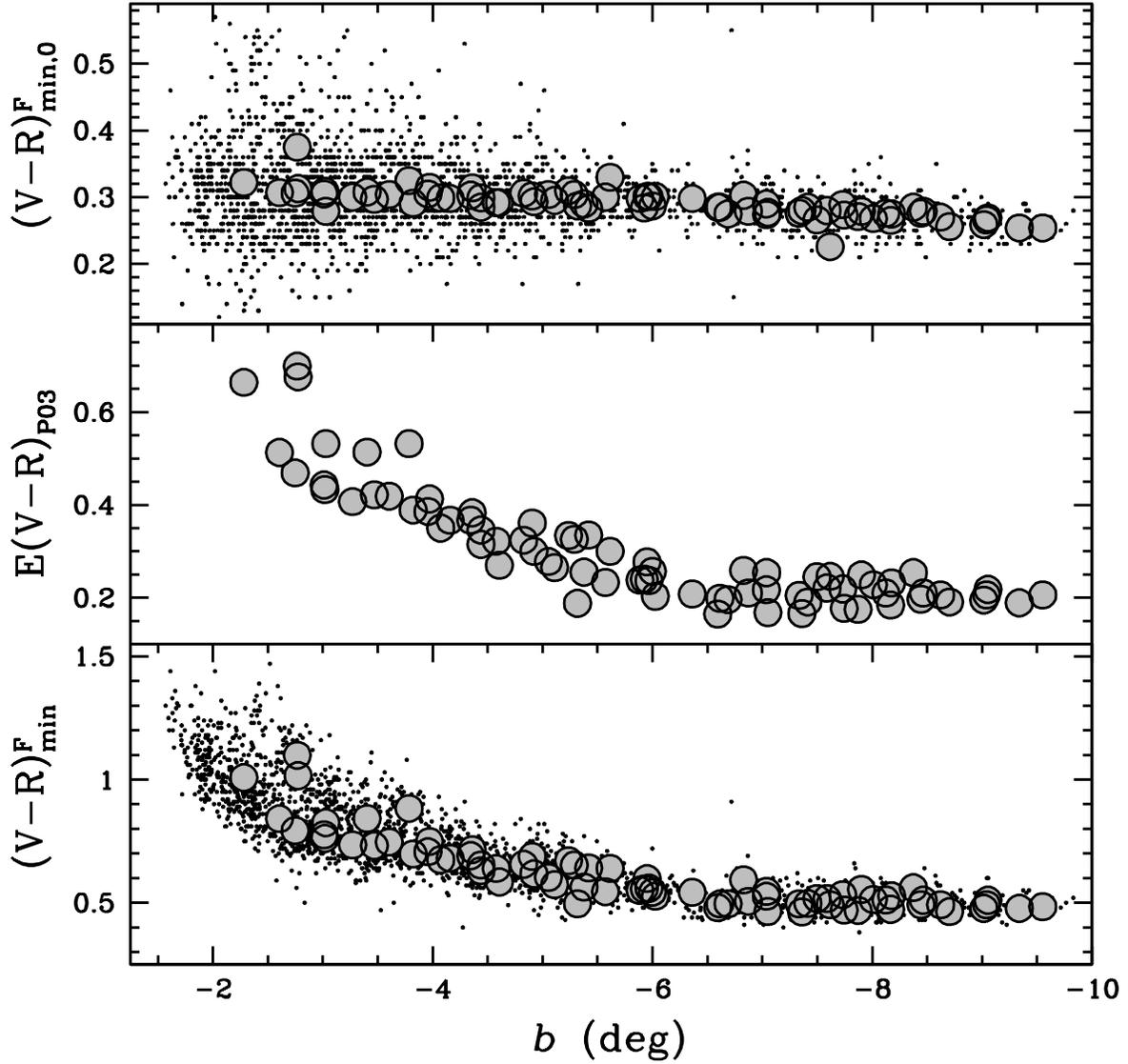}
\caption{
{\it bottom}:  The binned $\rm (V-R)$ color
at minimum $V$-band light of each MACHO field
as a function of Galactic $b$.  The individual
minimum light colors of the RR Lyrae
are represented by the small black dots.
{\it middle}:  The average color excess, $\rm E(V-R)$, from P03
binned by MACHO field as a function of Galactic $b$. 
{\it top}: The average intrinsic color at minimum 
$V$-band light of each MACHO field, dereddened according 
to P03, as a function of Galactic $b$.  
\label{fig7}}
\end{figure}

In the bottom panel of Figure~\ref{fig7}, the average 
$\rm (V-R)$ color at minimum $V$-band light, $\rm (V-R)_{min}^F$ 
of each MACHO field is plotted as a function of $b$.  The 
middle panel shows the average color excess as determined 
by Popowski, Cook \& Becker (2003) (hereafter P03),
$\rm E(V-R)_{P03}$, as a function of Galactic $b$.  The P03 
color excess values are based on the mean colors of stars 
from the MACHO survey toward the Galactic bulge, 
$\rm <V-R>$, in 4$^\prime$ x 4$^\prime$ regions of the sky.  
They show that $\rm <V-R>$ can be converted to extinction and 
visual extinction for 9717 elements at a resolution of about 
4$^\prime$ is determined.  The conversion from the P03 $\rm <V-R>$ 
to $\rm E(V-R)_{P03}$ used here is based on an exponential 
dusty disk model of the Galactic disk which is shown to 
correspond well with the $A_V$ reddening map from 
\citet{stanek96}.  This in contrast to using the conversion 
based on \citet{dutra03} reddening map, derived from Two 
Micron All Sky Survey data in the $J$, $H$, and $K$ bands.  
The \citet{stanek96} reddening map is not only in a passband 
that does not require additional conversions for the purposes 
used for here, but also has a zero point that has been accurately 
measured with two independent types of stars (Alcock et~al. 1998, 
Gould, Popowski \& Turndrup 1998).

The top panel of Figure~\ref{fig7} shows the intrinsic $\rm (V-R)$ 
color at minimum $V$-band light, $\rm (V-R)_{min,0}^F$ as a function of $b$.
There is a clear trend between $\rm (V-R)_{min,0}^F$ and Galactic 
$b$, with a slope of 0.008 $\pm$ 0.0008 mag/$^\circ$.  This plot illustrates 
the difficulty of making an accurate analysis 
using stars in the Bulge.  \citet{kundchab08} use 2435 RR0 Lyrae stars 
from the MACHO Galactic bulge fields to investigate the structure of the 
Galactic bulge.  They find no significant trends with metallicity, period 
or amplitude as a function of position.  Hence, there is no physical reason 
why $\rm (V-R)_{min,0}^F$ would be a function Galactic $b$.  

The chances of an RR Lyrae being blended increases closer to the Galactic 
plane, but this does not appear to be a contributing factor in the
$\rm (V-R)_{min,0}^F$ - Galactic $b$ trend.  \citet{popowski05} found 
evidence for blending for 1 event out of 53, ($\rm \sim2\%$) at $b$ $\rm >-$4
in an analysis of MACHO red clump stars to determine to what 
degree the clump microlensing events are blended.  
As clump giants have an average absolute $V$ magnitude that is 
$\sim$0.5 mag fainter than RR Lyrae stars, we would expect blending to be 
even less of a problem for the RR Lyrae.  

The P03 reddening values are based on the CMD of 4$^\prime$ x 4$^\prime$ regions
of the sky.  However, it is now known that there is a change in the CMD 
toward the plane.  From $\sim$700 giant stars, \citet{zoccali08} found that 
there is a metallicity gradient of $-$0.25 dex from $b$ = $\rm -12^\circ$ 
to $b$ = $\rm-4^\circ$.  Using isochrones, we find that a 0.25 dex change in 
metallicity corresponds to a change in color on the RGB of 0.06 mag.  The
average difference in $\rm E(V-R)$ between the RR Lyrae and P03 
from $b$ = $\rm -10^\circ$ to $b$ = $\rm-4^\circ$ is $\sim$$-$0.04 mag,
and extrapolating this result to cover $b$ = $\rm -12^\circ$ to 
$b$ = $\rm-4^\circ$, the difference in color excess is $-$0.053 mag.

The individual RR Lyrae $\rm (V-R)_{min,0}^F$ values are over-plotted
in the top and bottom panels Figure~\ref{fig7}.  It is clear that the 
dispersion in the RR Lyrae colors increases closer to the Galactic 
plane (lower $\rm|$$b$$\rm|$ values).  
The closer to the plane, the larger the differential reddening in
one of the P03 4$^\prime$ x 4$^\prime$ area.  Since the individual RR Lyrae 
fall at random places within each of P03 reddening map pixel, it 
is expected that the dispersion of the individual RR Lyrae 
(top and bottom panels) would increase.  As blended RR Lyrae
would appear redder, this would also explain why 
there are some ultra-blue RR Lyrae as well as very red ones. 
We hence caution the reader that the P03 reddenings have
a Galactic $b$ dependence and can underestimate the amount
of color excess, $\rm E(V-R)$, in low Galactic $b$ regions 
by $\sim$0.04 mag.

\begin{figure}[htb]
\includegraphics[width=16cm]{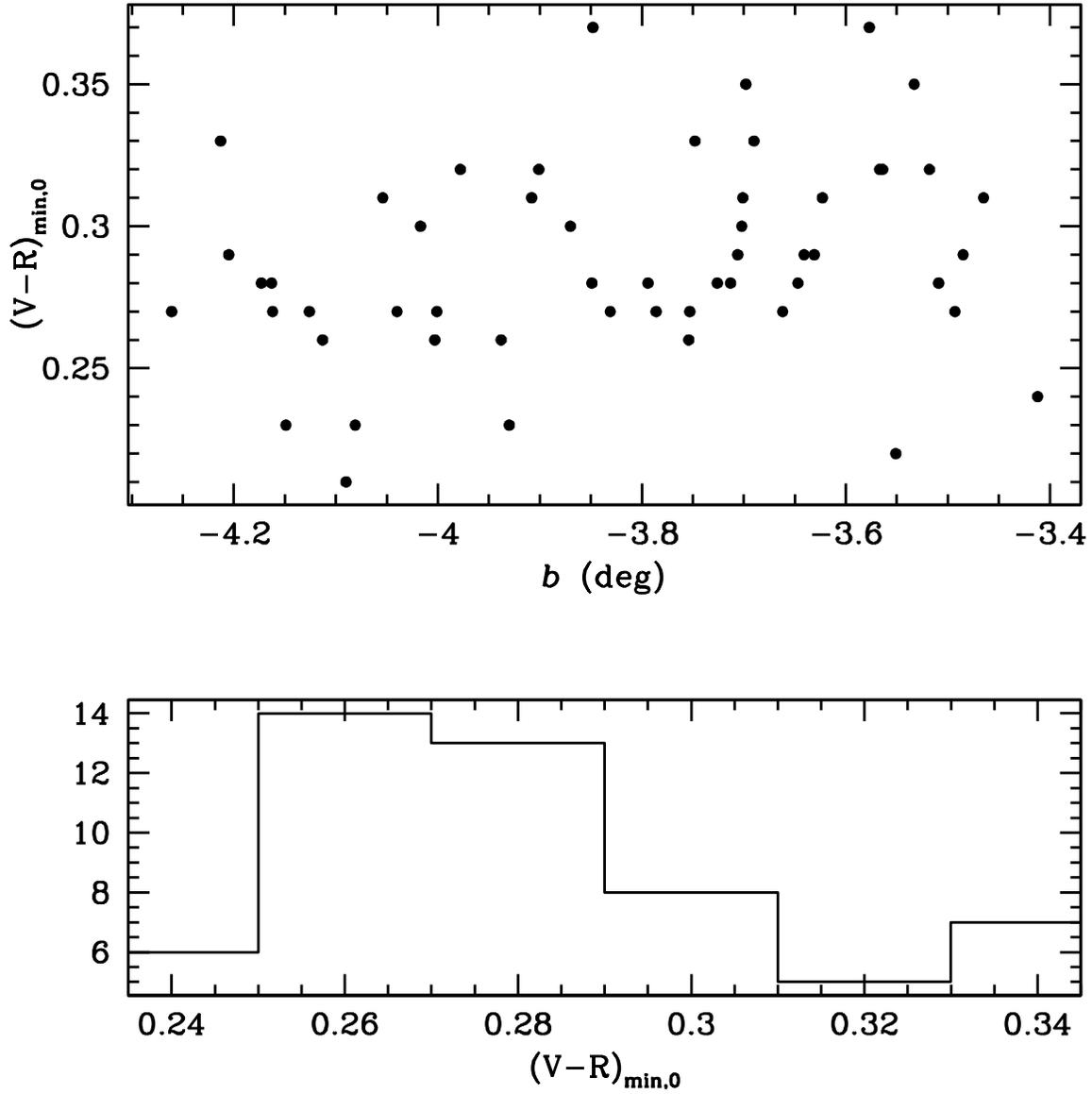}
\caption{{\it top}:  The 53 MACHO RR0 Lyrae stars in  Baade's 
Window with well determined minimum light colors, dereddened
according to P03, and plotted against Galactic $b$.
{\it bottom}:  The histogram of $\rm (V-R)_{min,0}$ for 53 BW
MACHO RR0 Lyrae stars, binned in 0.02 mag bins.
\label{fig8}}
\end{figure}

Baade's window centered is roughly on 
the globular cluster NGC 6522 at $(l, b) = (1 \hbox{$.\!\!^\circ$} 0,
-3 \hbox{$.\!\!^\circ$} 9)$ and is a region known to
have relatively low amounts of interstellar extinction.
We turn our analysis to this region.
In the top plot of Figure~\ref{fig8}, the 53 RR0 Lyrae stars in Baade's 
Window are shown as a function of Galactic $b$.  The histogram
of $\rm (V-R)_{min,0}$ values are shown in the bottom plot.  The median
of the distribution is 0.28 identical to that of the $\rm (V-R)_{min,0}$ 
found by the local RR Lyrae stars.  Their average $\rm (V-R)_{min,0}$
is 0.288 mag, and the mode is 0.27 mag.

\begin{figure}[htb]
\includegraphics[width=16cm]{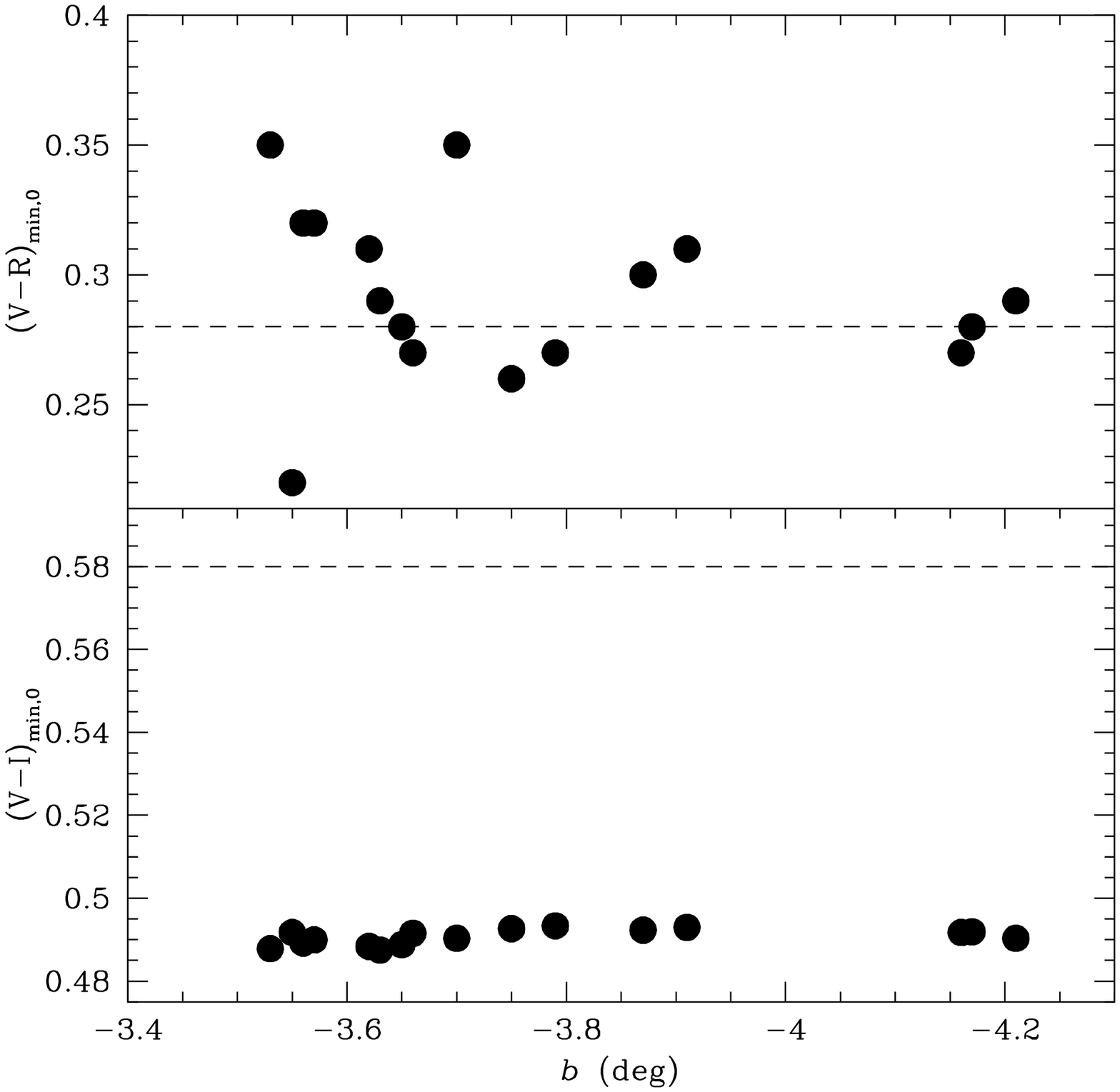}
\caption{{\it top}: Sixteen MACHO Bulge RR0 Lyrae star
$\rm (V-R)_{min,0}$ values, dereddened according to
P03.  The dotted line represents
the value of the local $\rm (V-R)_{min,0}$ value, determined
in this paper.
{\it bottom}:  The same 16 OGLE Bulge RR0 Lyrae star
$\rm (V-I)_{min,0}$ values, dereddened according to
\citet{sumi04}.  The dotted line represents
the value of the local $\rm (V-I)_{min,0}$ value, determined
by \citet{guldenschuh05}.
\label{fig9}}
\end{figure}

There are 17 RR0 Lyrae stars in Baade's Window that have OGLE 
$\rm (V-I)_{min,0}$ values from \citet{collinge06} that can be matched 
with a MACHO RR0 Lyrae star.  In Figure~\ref{fig9} these stars are shown 
as a function of Galactic $b$.  The dotted line indicates the local RR Lyrae
 $\rm (V-R)_{min,0}$ and $\rm (V-I)_{min,0}$ value, from the 
 $\rm (V-R)_{min,0}$ found above and from the \citet{guldenschuh05}
value, respectively.  The Bulge RR0 Lyrae $\rm (V-R)_{min,0}$ values
are similar to the local RR0 Lyrae value.  This is not the case for the
intrinsic Bulge RR Lyrae minimum light colors and local RR Lyrae 
$\rm (V-I)_{min,0}$ colors.  

From the above analysis, the $\rm (V-R)_{min,0}$ colors of the 
bulge RR Lyrae stars behave in a similar manner as local stars.  
Therefore we find it is unlikely that the $\rm (V-I)_{min,0}$ colors 
of the bulge RR Lyrae behave in an anomalous way from the local 
RR Lyrae stars.  This indicates that there either is a problem with 
the zero-point accuracy of the \citet{sumi04} reddening map, or 
with the \citet{guldenschuh05} $\rm (V-I)_{min,0}$ calibration.  
Given the difficulty of determing reddening in the Galactic bulge, 
we argue the most likely cause of this is a RR Lyrae-to-red clump 
color differential of the bulge population (measured from the OGLE data).  
Since we find the Bulge RR Lyrae stars do not exhibit anomalous colors,
the most likely explanation seems to be the influence of metallicity 
on the color of the red clump.  Unfortunately, systematic errors in the OGLE 
photometry may appear at approximately the level as 
the 0.05 mag discrepancy between the mean value of $\rm (V-I)_{min,0}$ 
in the OGLE Bulge and the local RR Lyrae sample \citet{collinge06}.
Hence, the dependence of the color of the red clump on metallicity is
complicated and would require a thorough look at the OGLE
photometry, which is beyond the scope of this paper.  

\section{Conclusion}
New $R$-band photometry for 21 local RR0 Lyrae stars is presented.  The
$R$-band light curves have between 65 - 145 data points per star.
Seventeen of these light curves are combined with $V$-band light curves from the 
literature, and the $\rm (V-R)$ color at minimum light
is found.  Fourteen additional stars from the literature are added to 
our sample.

Two definitions of $\rm (V-R)_{min}$ are explored.   $\rm (V-R)_{min}^F$ is 
defined as the minimum light color at minimum $V$-band light found 
by performing a Fourier decomposition to the $V$ and $R$ light curves.
$\rm (V-R)_{min}^{\phi(0.5-0.8)}$ is defined as the average color over
the phase range 0.5-0.8.  As our program stars have well known reddening 
values, the $\rm (V-R)_{min}$ colors are dereddened.  The
average dereddened color at minimum $V$-band light, $\rm (V-R)_{min,0}^F$,
is 0.28 $\pm$ 0.02, where 0.02 is the dispersion about the mean.  
The average dereddened color over the phase range 0.5-0.8, 
$\rm (V-R)_{min,0}^{\phi(0.5-0.8)}$, is 0.27 $\pm$ 0.02.  These values are 
based on a total sample of 31 stars, and do not appear to depend on \feh,
period, $V$-amplitude, or \dellpc.  The \feh range of our program stars 
spans \feh= 0 dex to \feh= $-$2.5 dex, the period range spans
P = 0.38 days to P=0.78 days, and the $V$-amplitude range spans 
$A_V$ = 0.39 mag to $A_V$ = 1.35 mag.  Previous hints of $V$-amplitude 
dependence on $\rm (V-R)_{min,0}^{\phi(0.5-0.8)}$ seem dispelled with the 
larger sample size.

Using the MACHO Bulge RR0 Lyrae stars from \citet{kundchab08},
we investigate if the Bulge RR0 Lyrae stars have anomalous
colors as compared to the local RR Lyrae stars.  The Bulge RR Lyrae
stars are dereddened using the \citet{popowski03} reddening map.
The RR0 Lyrae stars in Baade's Window have approximately the
same intrinsic $(V-R)_{min}$ as the local sample.  Hence the 
$\rm (V-R)_{min,0}$ colors of the bulge RR Lyrae stars behave 
in the same way as the $\rm (V-R)_{min,0}$ colors of local stars.
Given previous suggestions of anomalous bulge RR Lyrae colors
and the wide use of RR Lyrae variables as distance indicators,
such a convergence of results is particularly appreciated.

\clearpage

\renewcommand{\thetable}{\arabic{table}}
\setcounter{table}{0}  

\clearpage

\begin{table}
\begin{scriptsize}
\centering
\caption{Photometric Transformations}
\label{tab1}
\begin{tabular}{lcccccccc} \\ \hline
Night & JD & $N_{flds}$ & $rms_R$ & $N_R$ & $rms_V$ & $N_V$ & $rms_{R,nc}$  \\ \hline
May 6, 2006 & 2453861 & 6 & 0.017 & 38 & 0.011 & 40 & -- \\
May 7, 2006 & 2453862 & 7 & 0.023 & 50 & 0.017 & 47 & --\\
May 8, 2006 & 2453863 & 6 & 0.025 & 37 & 0.018 & 32 & 0.025 \\
\hline
\end{tabular}
\end{scriptsize}
\end{table}

\begin{table}
\begin{scriptsize}
\centering
\caption{Photometric parameters of program RR Lyrae stars}
\label{RRmags}
\begin{tabular}{lcccccccc} \\ \hline
Star & Period(d) & $<R>$ & $A_R$ & $N_{obs}$ & $F_{R,order} $ & $N_{VR}$ & $N_{R}$\\ \hline
$^1$UY Boo & 0.6508365 & 10.76 & 0.52 & 70 & $F_8$ & 4 & 4\\ 
TV CrB & 0.5846145 & 11.74 & 0.98 & 78 & $F_8$ & 2 & 3 \\ 
$^2$SS CVn & 0.47853 & 11.73 & 0.95 & 36 & $F_8$ & 3 & 0 \\ 
UZ CVn & 0.69779191& 11.90 & 0.74 & 105 & $F_8$ & 5 & 4  \\ 
BC Dra & 0.719576 & 11.27 & 0.47 & 129 & $F_6$ & 1 & 4  \\
BT Dra & 0.588673 & 11.42 & 0.64 & 134 & $F_8$ & 2 & 2 \\
SW Dra & 0.56966993 & 10.29 & 0.74 & 115 & $F_8$ & 2 & 0 \\
CW Her & 0.6238405 & 12.41 & 0.87 & 85 & $F_8$ & 2 & 3  \\.
VZ Her & 0.44032789 & 11.36 & 1.05 & 70 & $F_8$ & 1 & 8 \\
$^3$AX Leo & 0.726776 & 12.02 & 0.46 & 114 & $F_6$ & 2 & 0 \\
FN Lyr & 0.5273984 & 12.59 & 1.00 & 92 & $F_6$ & 2 & 4  \\
IO Lyr & 0.57712215 & 11.59 & 0.78 & 81 & $F_8$ & 2 & 4 \\
V413 Oph & 0.44900586 & 11.70 & 0.95 & 76 & $F_8$ & 2 & 1 \\
AN Ser & 0.52207162 & 10.73 & 0.79 & 96 & $F_8$ & 2 & 4  \\
AT Ser  & 0.746568 & 11.26 & 0.70 & 88 & $F_6$ & 2 & 3  \\
AW Ser & 0.597114344 & 12.69 & 0.97 & 86 & $F_8$ & 2 & 4 \\
$^4$AB UMa & 0.599577 & 10.74 & 0.36 & 47 & $F_6$ & 4 & 5 \\
$^5$RV UMa & 0.468060 & 10.64 & 0.58 & 84 & $F_8$ & 5 & 3 \\
AT Vir & 0.5257931 & 11.18 & 0.95 & 64 & $F_8$ & 4 & 3  \\
AV Vir & 0.656909 & 11.56 & 0.62 & 84 & $F_8$ & 4 & 2 \\
ST Vir & 0.410820 & 11.41 & 0.99 & 85 & $F_8$ & 2 & 3 \\
\hline

$^1$HJD 4533.7143 - 4541.9318 \\
$^2$HJD 4538.7370 - 4543.9270 \\
$^3$HJD 4533.6980 - 4543.8652 \\
$^4$HJD 4533.8675 - 4536.8210 \\
$^5$HJD 4533.7283 - 4543.9340 \\
\hline
\end{tabular}
\end{scriptsize}
\end{table}

\begin{scriptsize}
\begin{sidewaystable}[p]\small
\caption{Minimum light colors at minimum $V$-band light for field RR0 Lyrae stars}
\label{redval}
\begin{tabular}{lccccccccccc} \\ \hline
Star & $\rm E(B-V)$ & $\rm E(V-R)$ & $\rm A_V$ & $\rm [Fe/H]$ & \dellpc & $\rm (V-R)_{min,0}^F$ & $\rm \sigma (V-R)_{min}^F$ & $\rm F_{V,order} $ & $\rm (V-R)_{min,0}^{\phi(0.5-0.8)}$ & $\rm N_{min,V}$ & $\rm N_{min,R}$ \\ \hline
TZ Aur & 0.064 & 0.051 & 1.33 & $-$0.79 & 0.18 & 0.25 & 0.01 & $F_8$ & 0.238 $\pm$ 0.005 & 227 & 227\\
UY Boo & 0.037 & 0.030 & 0.79 & $-$2.56 & 0.03 & 0.28 & 0.02 & $F_4$ & 0.295 $\pm$ 0.018 & 15 & 4 \\ 
UY Boo & 0.037 & 0.030 & 0.79 & $-$2.56 & 0.03 & 0.30 & 0.02 & $F_8$ & 0.300 $\pm$ 0.011 & 16 & 24 \\ 
AL CMi & 0.030 & 0.024 & 0.72 & $-$0.79 & 0.15 & 0.32 & 0.01 & $F_8$ & 0.323 $\pm$ 0.031 & 13 & 13\\
TV CrB & 0.040 & 0.032 & 1.34 & $-$2.33 & 0.03 & 0.28 & 0.015 & $F_8$ & 0.251 $\pm$ 0.006 & 60 & 12\\
UZ CVn & 0.024 & 0.019 & 0.85 & $-$1.89 & $-$0.01 & 0.28 & 0.01 & $F_4$ & 0.282 $\pm$ 0.025 & 4 & 20\\
BC Dra & 0.069 & 0.055 & 0.59 & $-$2.00 & 0.01 & 0.26 & 0.01 & $F_8$ & 0.273 $\pm$ 0.030 & 16 & 28\\
BT Dra & 0.010 & 0.008 & 0.78 & $-$1.75 & 0.10 & 0.25 & 0.015 & $F_{8}$ & 0.241 $\pm$ 0.007 & 42 & 42\\
SW Dra & 0.015 & 0.012 & 0.91 & $-$1.12 & 0.09 & 0.27 & 0.02 & $F_8$ & 0.263 $\pm$ 0.015 & 17 & 26\\
CW Her & 0.022 & 0.018 & 1.07 & $-$2.03 & 0.00 & 0.25 & 0.01 & $F_4$ & 0.272 $\pm$ 0.007 & 5 & 23\\
VZ Her &  0.031 & 0.025 & 1.31 & $-$1.02 & 0.13 & 0.27  & 0.03 & $F_8$ & 0.283 $\pm$ 0.007 & 9 & 17\\
GO Hya & 0.048 & 0.038 & 0.54 & $-$0.77 & 0.10 & 0.33 & 0.01 & $F_4$ & 0.251 $\pm$ 0.030 & 5 & 5\\
AX Leo & 0.037 & 0.030 & 0.63 & $-$1.72 & $-$0.01 & 0.28 & 0.01 & $F_4$ & 0.282 $\pm$ 0.046 & 3 & 3\\
IO Lyr &  0.067 & 0.054 & 0.98 & $-$1.14 & 0.07 & 0.33 & 0.01 & $F_{8}$ & 0.235 $\pm$ 0.027 & 2 & 17\\
AN Ser & 0.042 & 0.034 & 1.01 & $-$0.07 & 0.12 & 0.29 & 0.01 & $F_8$ & 0.273 $\pm$ 0.030 & 36 & 32\\
AT Ser & 0.039 & 0.031 & 0.87 & $-$2.03 & $-$0.08 & 0.30 & 0.01 & $F_8$ & 0.266 $\pm$ 0.025 & 68 & 25\\
AW Ser & 0.039 & 0.031 & 1.20 & $-$1.67 & 0.00 & 0.25 & 0.01 & $F_8$ & 0.257 $\pm$ 0.011 & 36 & 21\\
AB UMa & 0.027 & 0.022 & 0.39 & $-$0.49 & 0.17 & 0.28 & 0.02 & $F_4$ & 0.249 $\pm$ 0.032 & 3 & 3\\
AB UMa & 0.027 & 0.022 & 0.39 & $-$0.49 & 0.17 & 0.29 & 0.02 & $F_4$ & 0.305 $\pm$ 0.031 & 4 & 4\\
AT Vir & 0.031 & 0.025 & 1.14 & $-$1.60 & 0.08 & 0.29 & 0.01 & $F_8$ & 0.278 $\pm$ 0.010 & 35 & 8\\
AV Vir & 0.032 & 0.026 & 0.72 & $-$1.25 & 0.04 & 0.29 & 0.01 & $F_8$ & 0.260 $\pm$ 0.017 & 47 & 24\\
ST Vir &  0.041 & 0.033 & 1.31 & $-$0.67 & 0.16 & 0.29 & 0.01 & $F_8$ & 0.256 $\pm$ 0.015 & 48 & 20\\
\hline
\end{tabular}
\end{sidewaystable}
\end{scriptsize}

\clearpage

\clearpage 

\clearpage

\clearpage

\begin{center}
    {\bf APPENDIX}
\end{center}

\renewcommand{\thetable}{A-\arabic{table}}
\setcounter{table}{0}  

\appendix
\section{$R$ Photometry of Program Stars}

The data tables can be found in the electronic version of the paper.

\end{document}